# More Precise Methods for National Research Citation Impact Comparisons[1]

Ruth Fairclough, Mike Thelwall
Statistical Cybermetrics Research Group, School of Mathematics and Computer Science, University of Wolverhampton, Wulfruna Street, Wolverhampton WV1 1LY, UK.
Tel. +44 1902 321470. Fax +44 1902 321478. Email: m.thelwall@wlv.ac.uk

Governments sometimes need to analyse sets of research papers within a field in order to monitor progress, assess the effect of recent policy changes, or identify areas of excellence. They may compare the average citation impacts of the papers by dividing them by the world average for the field and year. Since citation data is highly skewed, however, simple averages may be too imprecise to robustly identify differences within, rather than across, fields. In response, this article introduces two new methods to identify national differences in average citation impact, one based on linear modelling for normalised data and the other using the geometric mean. Results from a sample of 26 Scopus fields between 2009-2015 show that geometric means are the most precise and so are recommended for smaller sample sizes, such as for individual fields. The regression method has the advantage of distinguishing between national contributions to internationally collaborative articles, but has substantially wider confidence intervals than the geometric mean, undermining its value for any except the largest sample sizes.
**Keywords**: scientometrics; citation analysis; research evaluation

## 1. Introduction

The task of monitoring or evaluating large groups of researchers is driven by the need to justify funding and to assess the effects of policy changes. At the national level, this may be undertaken by government departments or by others on their behalf. A standard approach is to compare the average citation impact of a country's outputs with those of other countries (Aksnes, Schneider, & Gunnarsson, 2012; Albarrán, Crespo, Ortuño, & Ruiz-Castillo, 2010; Albarrán, Perianes-Rodríguez, & Ruiz-Castillo, 2015; Elsevier, 2013; Jiménez-Contreras, de Moya Anegón, & López-Cózar, 2003; King, 2004). Individual fields (Schubert & Braun, 1986) or sets of fields (Braun, Glänzel, & Grupp, 1995; Ingwersen, 2000) may also be compared internationally, for example to identify areas of excellence. Nevertheless, citation data is highly skewed (de Solla Price, 1976), making conventional arithmetic mean impact estimates unreliable, especially when little data is available. Thus, methods that work reasonably well for comparing entire countries may not be precise enough to compare individual fields between countries because of the smaller number of publications involved. Hence, alternatives to comparisons of mean numbers of citations per paper may be needed for field-level comparisons.

Although international comparisons based on citation counts are relatively transparent and objective, they have unavoidable substantial biases in practice. The use of citation counts as an impact indicator is intrinsically problematic because articles can be cited for reasons unrelated to their academic value (MacRoberts & MacRoberts, 1989), even if, on a theoretical level, citations should perhaps be used mainly to acknowledge important prior work (Merton, 1973). On a large scale, however, unwanted types of citation may tend to even

---

out so that it is reasonable to compare the overall average citation counts (van Raan, 1998). Significant positive correlations between peer judgements and citation indicators are evidence of the value of this approach (Franceschet & Costantini, 2011; Gottfredson, 1978; HEFCE 2015), but citation indicators should only be used to inform rather than replace human judgements of impact because of the variety of the reasons why research is valuable and why articles are cited. Perhaps most problematically, the coverage of the citation index used influences the results in unpredictable ways. Citation indexes do not have comprehensive coverage and the extent of coverage of national journals is likely to vary substantially (Van Leeuwen, Moed, Tijssen, Visser, & Van Raan, 2011). In particular, although Scopus seems to have wider coverage than the Web of Science (López-Illescas, de Moya-Anegón, & Moed, 2008), it indexes a lower proportion of non-English than English academic journals (de Moya-Anegón, et al., 2007). This could be an advantage for countries that publish poor quality research in their national non-English journals because the low cited articles in these will not be included in the citation average calculations. Conversely, however, if a nation's best publications are in national non-English journals then its citation average may suffer from their exclusion. Despite these limitations, citation-based international comparisons are widely used in the absence of viable alternatives or as one of a range of indicators (Elsevier, 2013).

In response to the need for more precise indicators for comparisons of international scholarly impact between fields, this article introduces two new methods that reduce the variation in citation data through normalisation. The first method is to use statistical modelling on transformed data in order to estimate the underlying geometric mean citation count for each country within a subject. The second method uses geometric means directly for each country, without any modelling. The geometric mean is based upon the arithmetic mean of the log of the data and is more suitable than the basic arithmetic mean for highly skewed data, such as citation counts, because it is less influenced by individual high values (Zitt, 2012). Geometric mean have been previously used for journal impact calculations (Thelwall & Fairclough, 2015a), but apparently not for international comparisons. Both methods should give more precise estimates than previous methods that have used non-normalised data and both methods allow relatively straightforward confidence interval estimates, without having to rely upon bootstrapping.

## 2. Research Questions

The objective of this study is to introduce, access and compare two new methods for national research impact indicators and to assess them for individual subjects. The following questions are motivated by this objective.
1. Do the new national subject-based citation impact estimation methods give comparable results to those of the previous standard methods for recent years?
2. Which of the new national subject-based citation impact estimation methods gives the results with the narrowest confidence intervals?
3. Are the new national subject-based citation impact estimation methods precise enough to reliably differentiate between major research nations for recent years within individual subjects?

## 3. Data and Methods

### *3.1 Data*

Lists of articles within defined fields from a specified set of recent years were needed to address the above questions. Recent years were used because impact comparisons are most

43policy relevant when applied to recent data and the use of multiple years allows trends over time to be identified. Scopus categories and Scopus data was chosen for this because Scopus has wider international coverage of journal articles than its main competitor (Van Leeuwen, Moed, Tijssen, Visser, & Van Raan, 2011). Although its subject categories are imperfect, they were chosen in preference to an alternative categorisation process using references or citations (Waltman & van Eck, 2012) to avoid the potential to bias the results by exploiting citations in the data selection phase. The following subject categories were chosen to represent a range of different subject areas: Animal Science and Zoology; Language and Linguistics; Biochemistry; Business and International Management; Catalysis; Electrochemistry; Computational Theory and Mathematics; Management Science and Operations Research; Computers in Earth Sciences; Finance; Fuel Technology; Automotive Engineering; Ecology; Immunology; Ceramics and Composites; Analysis; Anesthesiology and Pain Medicine; Biological Psychiatry; Assessment and Diagnosis; Pharmaceutical Science; Astronomy and Astrophysics; Clinical Psychology; Development; Food Animals; Orthodontics; Complementary and Manual Therapy. The Scopus data, including citations counts and author affiliation information, was downloaded from the Scopus from April 15 to May 11, 2015. Although it would be preferable to use a fixed citation window (e.g., count only citations within two years of publication for each article), this data was not available to the authors. The use of a variable citation window may affect all the indicators because, for example, highly cited articles might attract substantial numbers of citations over a long period of time, making them disproportionately influential for longer citation windows, even though in the analyses articles are only compared to other articles from the same year. Data was collected from each year from 2009 to 2015 to give a reasonable number of years for comparison. The partial year 2015 was included because policy makers are typically interested in the most current data possible and so it is useful to include the most recent year, even if it is unlikely to give useful results.

These datasets were re-used from a previous article (Thelwall & Fairclough, 2015b), for which they were employed within a combined set of 26 subjects to assess a new Mendeley-based method for identifying evidence of recent national differences in overall average citation impact but not individually by subject.

A set of countries was needed for the comparisons because it was not practical to compare all countries. The nine countries highlighted in a recent report were chosen (Elsevier, 2013) as a reasonable selection: USA; UK; Canada; Italy; Germany; France; China; Japan; Russia.

Each article was assigned a country proportion $p_c$ for each country equalling the proportion of article authors for that country, using the affiliation information given in Scopus and ignoring articles without author affiliations. This fractional counting method was chosen in preference to whole counting (allocating a full share of citations to each author), first author counting (allocating the whole article to the first author) and unequal counting (allocating different shares to each author based on their position in the authorship list). These are either problematic to justify in practice (unequal counting) or probably unfair in general (the others) (Aksnes, Schneider, & Gunnarsson, 2012; Huang, Lin, & Chen, 2011; Waltman & van Eck, 2015). In any case, authorship practices vary between and within disciplines, including alphabetical (Levitt & Thelwall, 2013; van Praag & van Praag, 2008), first author priority (Engers, Gans, Grant, & King, 1999), descending order of contribution (Bhandari, et al., 2004; Marusic, Bosnjak, & Jeroncic, 2011), or with the last author making the second most important contribution (Baerlocher, Newton, Gautam, Tomlinson, & Detsky, 2007).

The citation data was transformed for the regression by adding 1 and taking the natural log of the result. This is a standard normalisation technique for highly skewed data that is positive but contains zeros and follows a discretised lognormal distribution (Thelwall & Wilson, 2014b), such as citation counts (Thelwall & Wilson, 2014a). Although any positive number other than 1 could also be added to give the same effect and there is no intrinsic justification for the use of 1, it has the advantage of being the most straightforward choice and hence is the default for an analysis in the absence of a good reason for a different number.

## *3.2 National subject-based citation impact estimation methods*

Citation data is known to be highly skewed and hence not normally distributed but can be approximately normally distributed after a logarithmic transformation (Thelwall & Wilson, 2014a). To test this assumption, Normal Q-Q plots were checked for some years and subjects and skewness and kurtosis statistics calculated for each subject and year (182 combinations). The skewness and kurtosis values were outside of the acceptable range (-3 to +3 is a rule of thumb) for all years for the citation data but were in the acceptable range for all years until 2013 for the logarithmic citations. This suggests that the data is sufficiently close to the normal distribution in overall shape that calculations based upon assumptions about the normal distribution will give reasonably accurate results. With the logarithmic transformation, therefore, standard least squares regression and confidence interval formulae for the normal distribution can be used with confidence until 2013 and with some suspicion for 2014, but the values for 2015 cannot be trusted.

**Table 1**. Mean skewness and kurtosis values for each year (n=26 per year).

| Year | Citations skewness | Citations kurtosis | ln(1+citations) skewness | ln(1+citations) kurtosis |
|---|---|---|---|---|
| 2009 | 11.1 | 400.9 | 0.1 | 2.5 |
| 2010 | 10.2 | 366.8 | 0.1 | 2.5 |
| 2011 | 9.6 | 370.8 | 0.2 | 2.6 |
| 2012 | 5.9 | 105.7 | 0.4 | 2.6 |
| 2013 | 6.6 | 147.5 | 0.7 | 3.0 |
| 2014 | 7.0 | 167.3 | 1.5 | 4.9 |
| 2015 | 7.8 | 116.5 | 4.2 | 23.6 |

**Linear regression for the geometric mean:** For each subject and year, a statistical model was built to estimate the mean normalised citation count for articles from each country.

$$log\,(1\,+\,citations)\,=\,a\,+\,\sum_{c} \beta_c\, p_c$$

Here the sum is over all countries, $p_c$ is the proportion of authors from country $c$, and $log$ is the natural logarithm. The solution of the linear regression model will give the constant $a$ and the individual contribution rate $\beta_c$ of each country. The ordinary least squares method was used to fit the regression model.

      The raw $\beta_c$ values must be transferred back to citation means to give more intuitive results and this is achieved with the transformation $e^{a+\beta_c} - 1$ which is the expected geometric mean number of citations attracted by an article fully authored by country $c$, as predicted by the model. If all countries produced research with the same geometric mean then these values would all be approximately equal to the overall geometric mean, $\mu_g = e^{\frac{1}{n}\sum log(1+citations)} - 1$, where $n$ is the number of articles and the sum $\sum log(1 + citations)$ in the formula is over all articles. Thus the national bias estimated by the regression model is

therefore the expected geometric mean citations for country *c* divided by the geometric mean citation count for all articles:

$$(e^{a+\beta_c} - 1)/\mu_g \quad (1)$$

**Geometric mean comparisons:** A simpler approach is to calculate the mean of $log\,(1 + citations)$ for the articles from each country (i.e., the geometric mean of the citation counts, with an offset of 1), using weighting to reflect the author share in each article as follows.

$$\mu_{gc} = \exp\left(\frac{\sum_c log\,(1 + citations)p_c}{\sum_c p_c}\right) - 1$$

This formula gives a weighted geometric mean for each country. The bias for each country can again be obtained by dividing by the overall geometric mean:

$$\mu_{gc}/\mu_g \quad (2)$$

The second formula is almost identical to the first, since the denominator of both is the same and the numerator in both cases is a national geometric mean estimate. Formula (1) should better reflect the contribution to impact of a nation's research, however, because the linear model fitted can adjust for differing contributions to an article between different countries in their collaborative articles, whereas (2) assumes that all authors contribute equally. An example using arithmetic means for simplicity can illustrate this. Suppose that country A authors one solo paper with 12 citations and one joint paper with country B, with an author from each country, that has 6 citations. Suppose that country B authors one solo article with 0 citations. Using fractional counting, country A has a mean of (12+0.5x6)/1.5=10 citations per paper and country B has a mean of (0+0.5x6)/1.5=2 citations per paper, so A seems to be five times as productive as B. With the regression approach, the model fitted for citations per paper is $6 + 6p_A\text{-}6p_B$ and so papers from A expect to get 6 + 6x1 - 6x0=12 citations whereas papers from B expect to get 6 + 6x0 - 6x1=0 citations. Thus the 6 citations from the joint paper have been solely "contributed" by country A with none from country B, which seems reasonable in the light of the citations attracted by their solo articles.

**Arithmetic mean comparisons:** The previously used approach, which seems to be standard for bibliometrics in international comparison reports, is identical to (2) except with the arithmetic mean rather than the geometric mean, as follows.

$$\mu_c = \frac{\sum_c citations \times p_c}{\sum_c p_c}$$

The national bias for each country can then be obtained by dividing by the overall arithmetic mean:

$$\mu_c/\mu \quad (3)$$

This is essentially the method used by both the old and new crown indicators (Waltman, van Eck, van Leeuwen, Visser, & van Raan, 2011a,b) because it is applied here only to each subject area separately rather than to a combined set of articles from multiple subjects.

**Proportion in the most cited X% comparisons:** Another common way to compare national performances is to calculate the proportion of a country's articles found in the most cited X% of all articles, where X may be taken as 1, 10, or other values (e.g., Elsevier, 2013; Levitt, & Thelwall, 2009). For this calculation, articles above the X% threshold were counted as well as a proportion of the articles exactly at the X%, each counting a fractional value equal to the overall fraction of articles at the X% level that are needed to make up exactly X% (Waltman, & Schreiber, 2013). For example, in a set of 10 articles, if 3 tied for the 50% threshold and 4 were above the 40% threshold, then each article from a country at the 50% threshold would count as 1/3. This method is relatively simple but the choice of X can be arbitrary outside of a particular policy need. Percentile approaches such as this are recommended for research evaluations (Hicks, Wouters, Waltman, de Rijcke, & Rafols, 2015).

## *3.3 Confidence interval calculations*

The width of the 95% confidence interval for the mean will be used as a measure of the precision of each estimate. In some cases confidence intervals can be calculated using a standard formula but in all cases estimates can be obtained using bootstrapping (DiCiccio & Efron, 1996).

**Linear regression for the geometric mean:** The linear regression model produces standard errors for all the parameters estimated, and these can be used to calculate 95% confidence intervals for the mean. These estimates should be reasonable for years when the data is approximately normally distributed (2009 to 2013) and perhaps 2014 but the 2015 values are likely to be crude estimates. For this, the confidence intervals for the intercept were ignored and the intervals calculated from the slope coefficients only. For large sample sizes, the confidence interval would therefore be as follows (replacing 1.96 with the appropriate *t* value for moderate sample sizes):

$$(e^{a+\beta_c \pm 1.96 SE(\beta_c)} - 1)/\mu_g \qquad (1ci)$$

**Geometric mean comparisons:** The national authorship weighted sample standard deviation $s_c$ of the transformed data can be used to calculate confidence intervals as follows, where $n_c = \sum_c p_c$ is the weighted authorship sum for country $c$ (i.e., the sum of the fractional contributions of each country to the articles).

$$(\mu_{gc} \pm s_c/\sqrt{n_c})/\mu_g \qquad (2ci)$$

Confidence intervals cannot always be calculated for the arithmetic mean because highly skewed distributions, such as those fitting citation counts, may not have a theoretical population mean (Newman, 2005). Even though citation distributions are always finite and so can have a sample mean because there is a finite number of citing documents, if the theoretical distribution that they best fit has an infinite mean then it would be impossible to estimate that theoretical mean. Similarly, although bootstrapping can be used to estimate confidence intervals (DiCiccio & Efron, 1996) its results would be misleading in the absence of a theoretical population mean.

Most

**Proportion in the most cited X% comparisons:** For large samples, there is a standard formula to calculate confidence intervals for a sample proportion, as long as it is not zero or too close to zero (Sheskin, 2003). This formula assumes an infinite population and this is a problem for large countries. For example, a country that published half of the world's papers could get at most 2% of its papers in the world's top 1% (i.e., it published all of the world's top 1% research) whereas a country that published 1% of the world's papers could conceivably have all of them in the top 1%, so would have 100% of its papers in the top 1%. X% papers could not get all of their papers in the top X%. Moreover, the formula is not designed to deal with fractional contributions, as with academic authorship. Nevertheless, it may give a reasonable estimate of the accuracy of the most cited X% statistic. The 95% confidence interval formula is as follows, with $t_c$ being the proportion of articles from country c in the top X%.

$$t_c \pm 1.96/\sqrt{t_c(1-t_c)/n_c} \qquad (4ci)$$

## 4. Results

The four methods were applied to all 26 subjects. Detailed graphs for each individual subject are available online[2] as are the programs used for the statistical analyses[3] and the medians of the results are reported below (Figures 1 to 4). The "Others" category is not shown in Figure 1 because the variables are linearly dependant (they sum to 1) and so not all can be estimated with linear regression. The graphs show medians rather than means because some subject areas exhibited anomalous behaviour and the use of medians reduces the influence of individual unusual cases. The anomalous behaviour of Russia in Figures 1 to 3 for 2015 is made possible by the low number of Russian articles (a median of 9 articles per subject in 2015). The modelled geometric mean approach (formula 1; Figure 1) gives almost identical results to the simple geometric mean (formula 2; Figure 2) so the impact of different national contributions to collaborative articles has a minor impact on the overall results. Nevertheless, there are some consistent small differences. The most noticeable is the case of Russia, which has lower results for the model than for the geometric mean calculation. This suggests that Russian contributions to internationally collaborative research tend to be less successful at attracting citations than those of their partners. In other words, Russian researchers gain more from international collaboration leading to Scopus-indexed papers than do their international partners, at least in terms of Scopus-indexed citations.

The arithmetic mean results are broadly similar to the other two methods, but with noticeable differences, such as the reversed relative positions of Japan and Others between Figures 2 and 3. This gives some confidence that the new methods do not give strange results, which might undermine their value. The top 10% statistics (Figure 4) shows a tendency for convergence for recent years. This is a statistical artefact rather than a general trend, however. This is because with lower average citation counts for recent years it is more likely that articles with a tendency to receive few citations would be in the top 10%. Thus, only the first three formulae are realistic alternatives, assuming that the magnitude of differences between countries is of interest rather than just the rank order. This problem could be resolved by the use of a fixed citation window, however.

The scales of Figures 1 to 3 are all broadly comparable with each other. From Figure 3, for example, a typical paper in a typical subject from the USA in 2009 received 1.40 times as many citations as the world average for that subject. Similarly, from Figure 2, a typical paper in a typical subject from the USA in 2009 received 1.46 times as many citations as the world average for that subject. The difference is that "typical" in the first case refers to matching the arithmetic mean (and giving particularly high importance to individual highly cited papers) whereas the second typical article matches the geometric mean (and gives less importance to highly cited papers). The Figure 1 results also estimate the geometric mean and so have the same interpretation as Figure 2, except that for shared articles, Figure 1 allows different nationalities to contribute different amounts to articles, whereas Figure 2 does not.

---

[2] http://figshare.com/articles/Results_for_the_paper_More_Precise_Methods_for_National_Research_Citation_Impact_Comparisons/1544164

[3] http://figshare.com/articles/R_code_for_the_paper_More_Precise_Methods_for_National_Research_Citation_Impact_Comparisons/1544160
http://figshare.com/articles/R_code_for_the_paper_More_Precise_Methods_for_National_Research_Citation_Impact_Comparisons/1544159 and dummy data to try out the code with http://figshare.com/articles/Dummy_data_for_the_paper_More_Precise_Methods_for_National_Research_Citation_Impact_Comparisons/1544161

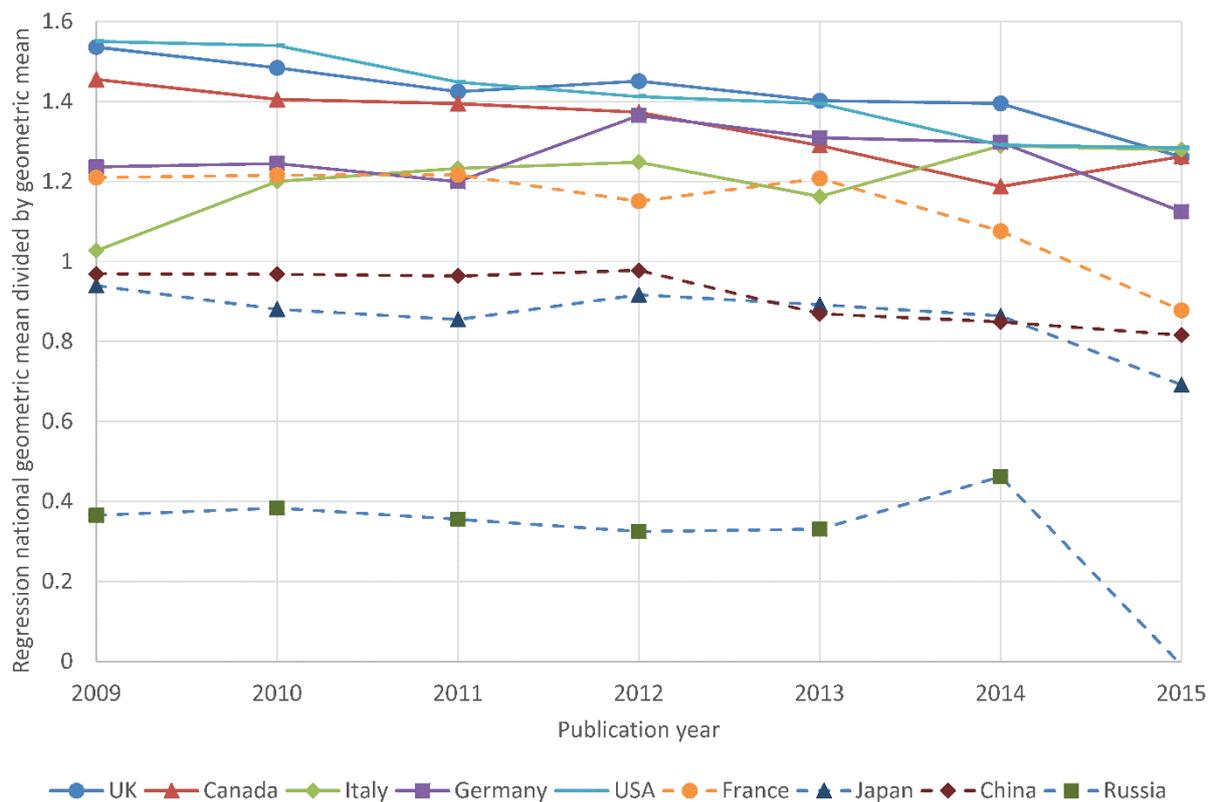

Figure 1. National geometric means for article citation counts estimated by the regression model, divided by the overall geometric mean (national citation impact estimator formula 1). Each point in the graph is the median across the 26 subjects.

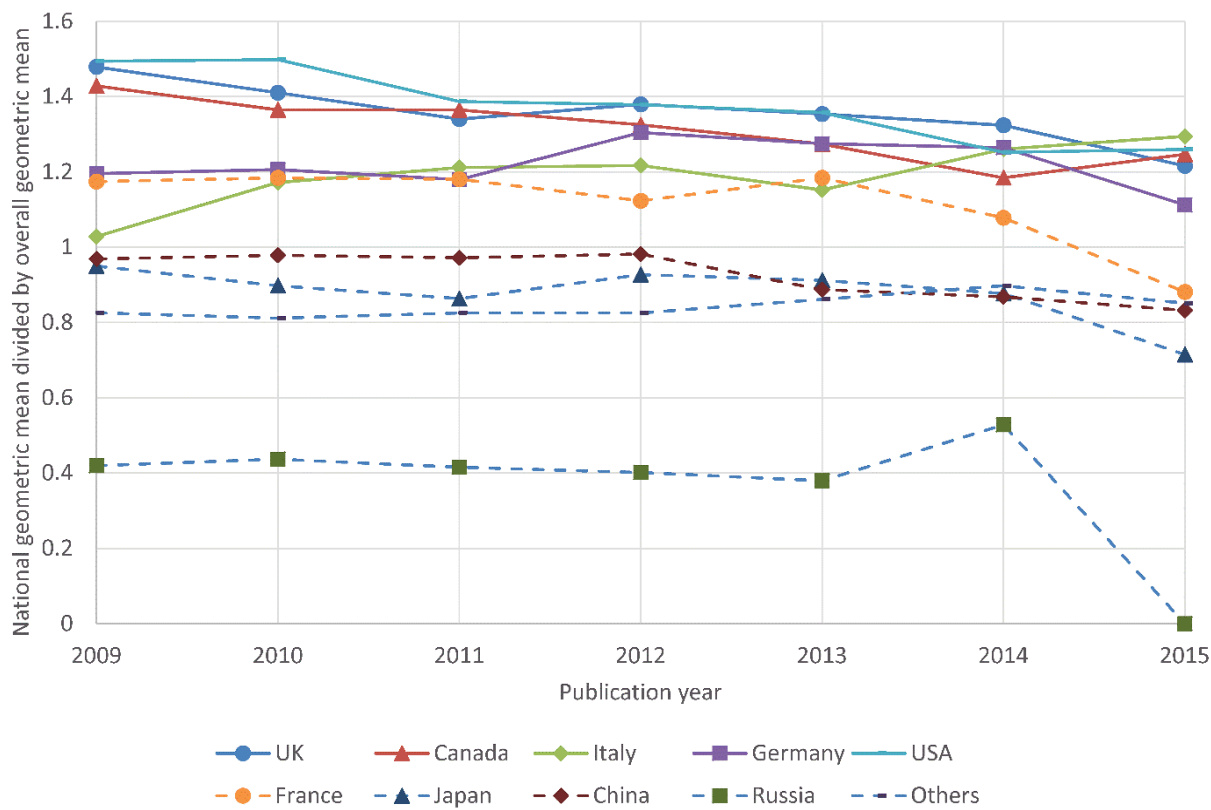

Figure 2. National geometric means for article citation counts divided by the overall geometric mean (national citation impact estimator formula 2). Each point in the graph is the median across the 26 subjects.

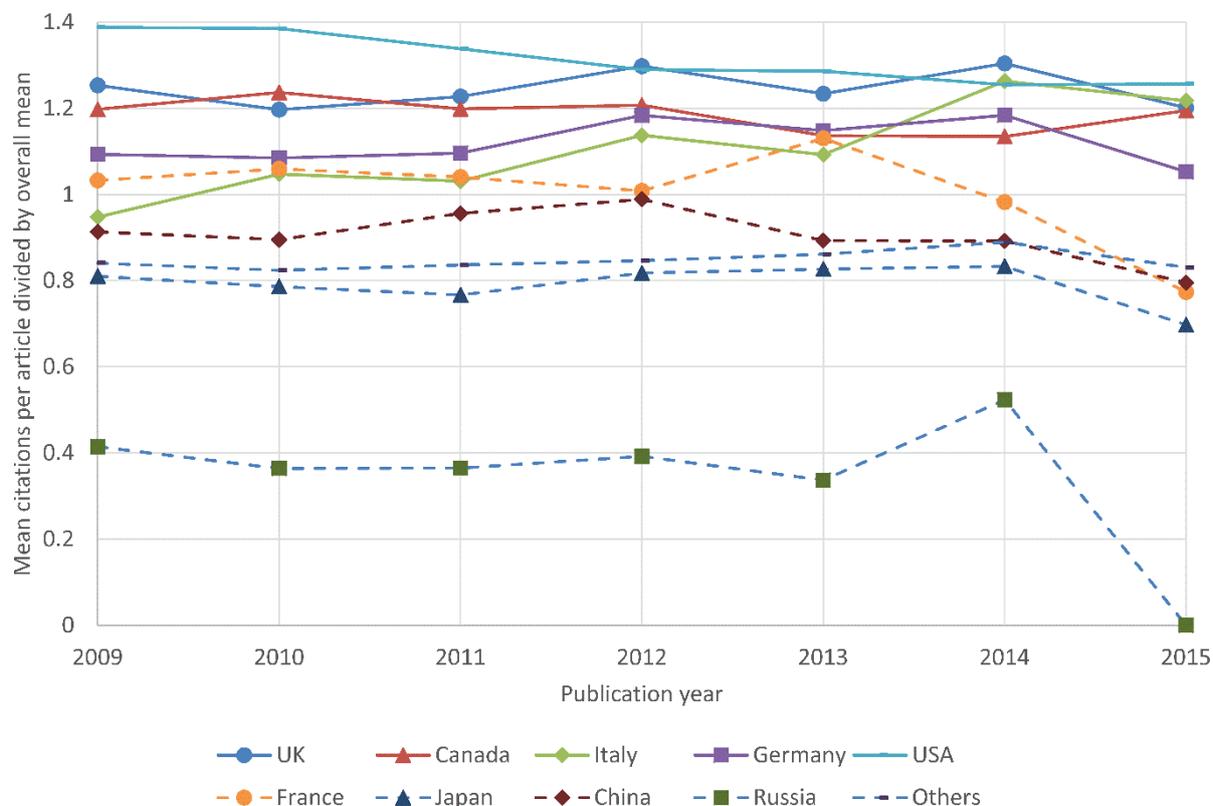

Figure 3. National arithmetic means for article citation counts divided by the overall arithmetic mean (national citation impact estimator formula 3). Each point in the graph is the median across the 26 subjects.

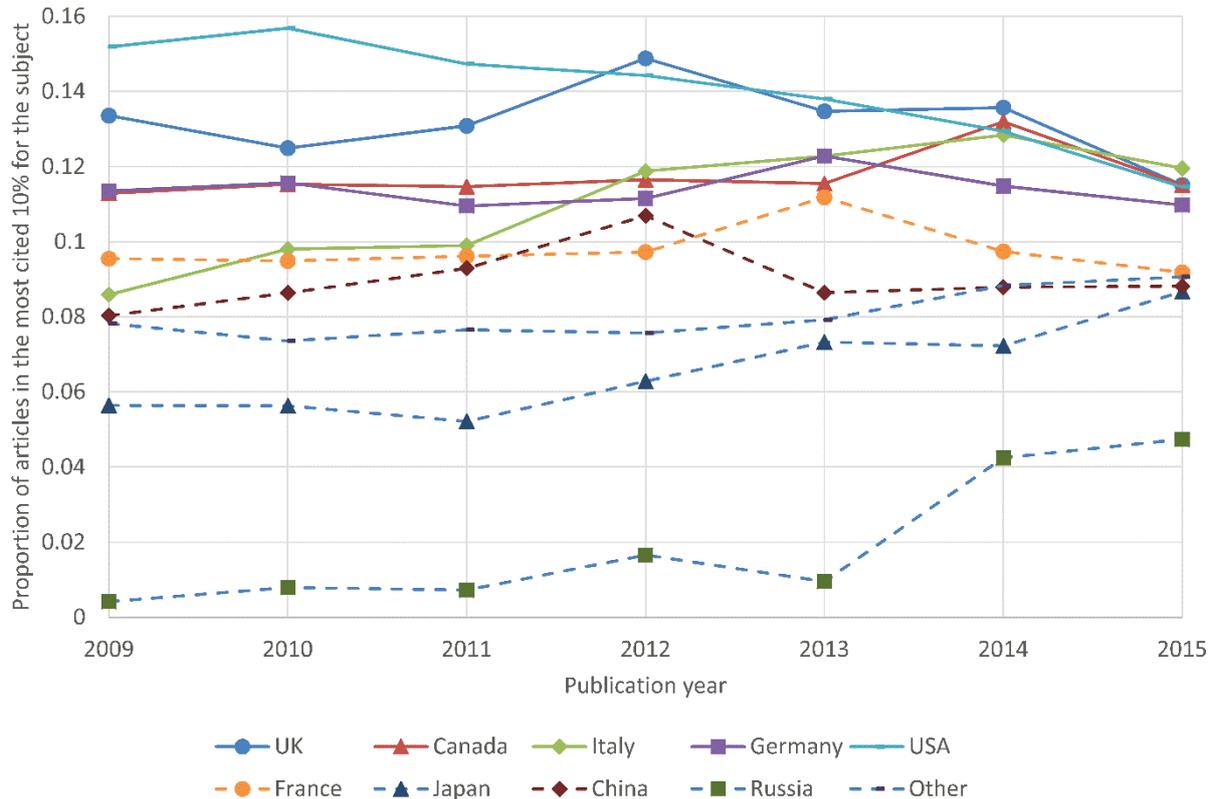

Figure 4. Proportion of articles from each country in the most highly cited 10%. Each point in the graph is the median of the values for the 26 subjects.

## *4.1 Stability*

Confidence intervals are a useful indicator of the accuracy of a parameter, such as the mean, that is estimated from a sample from a population. A 95% confidence interval, for example, suggests that if the calculation was to be continually repeated with new samples from the same population then 95% of the time the confidence interval would contain the correct population parameter. When the sample analysed is complete, such as all articles indexed by Scopus within a specific year and category, then a more abstract interpretation of the confidence interval can be made. In this case, the set of articles indexed in Scopus can be viewed as a sample from the theoretical population of articles that could be indexed by Scopus in similar circumstances. In practice, however, any calculations with real data for a confidence interval are reliant on a number of assumptions. For example, a simple formula can be used to calculate confidence limits for data that can be assumed to be normally distributed. The confidence intervals reported below should therefore be interpreted cautiously as indicators of the accuracy of the sample statistics.

The confidence intervals for the geometric means (Figure 6) are substantially narrower than those derived from the linear model (Figure 5), suggesting that the complexity of the fitting process for the linear regression model makes its predictions less precise than those of the geometric mean. Nevertheless, both are inadequate in the sense of being relatively large compared to the national impact estimates made and so it is not possible to draw any except the broadest statistical conclusions about the differences between nations for individual fields and subjects from the results. This is particularly true for the most recent year, 2015, but applies to all previous years as well. These conclusions are in spite of the relatively large sample sizes involved because the median number of articles for each subject and year is 2289 for 2015 and at least 5714 for each previous year.

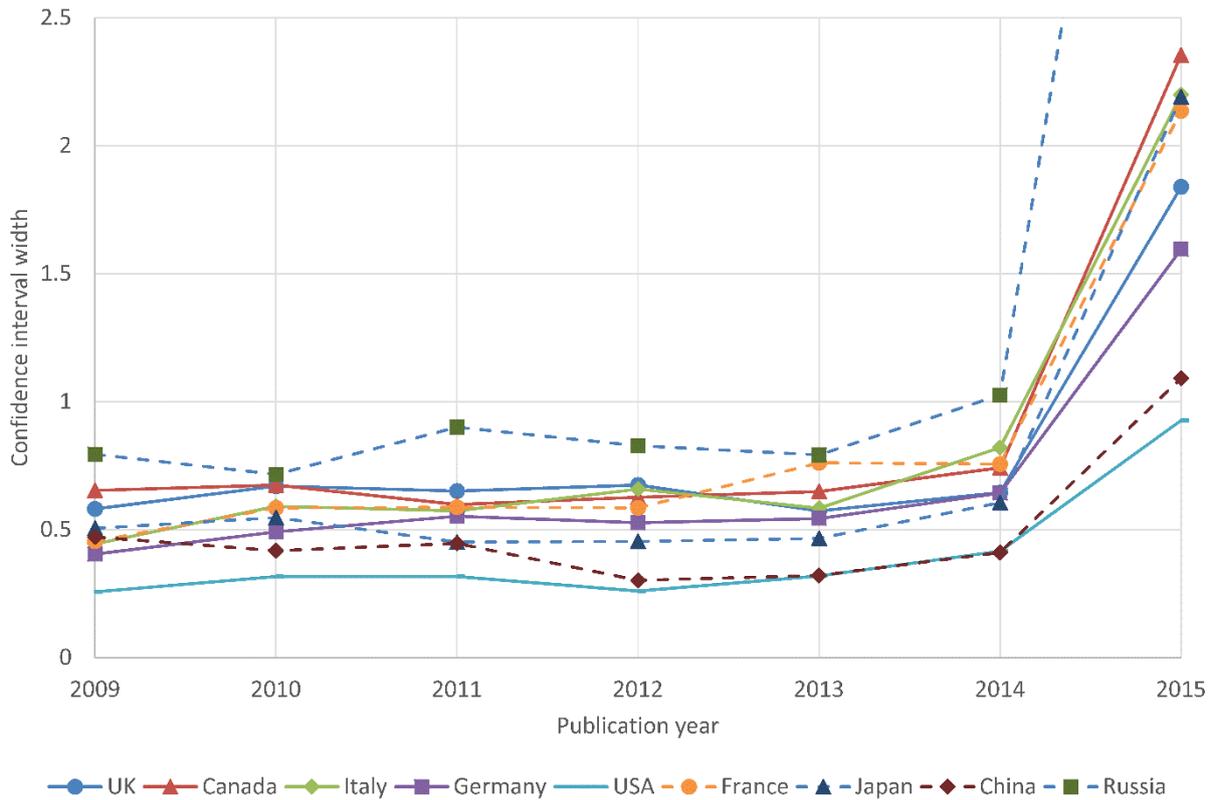

Figure 5. 95% confidence interval widths for the national citation impact estimators from the linear model (Figure 1). Each point in the graph is the median across the 26 subjects.

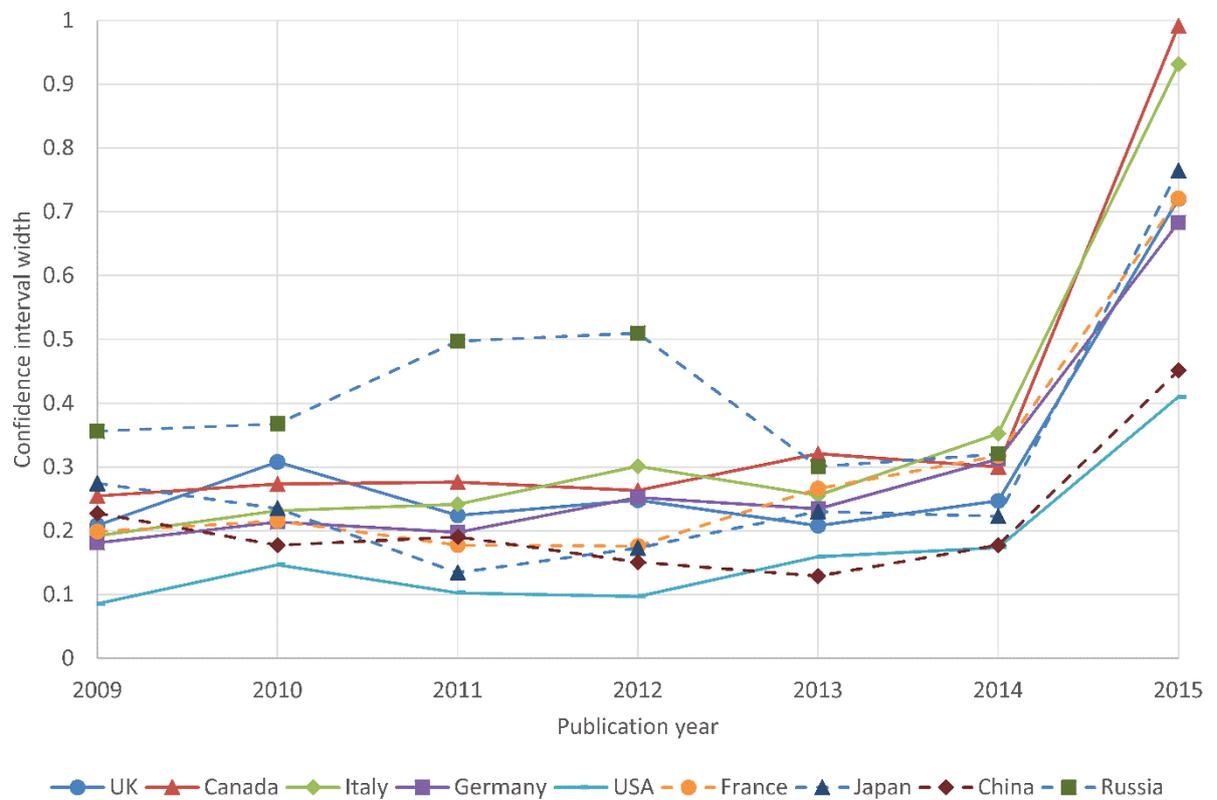

Figure 6. 95% confidence interval widths for the national citation impact estimators from the geometric means (Figure 2). Each point in the graph is the median across the 26 subjects.

Confidence intervals were also calculated with a standard bootstrap approach (resampling 999 times from the original sample and calculating the range in which 95% of the resulting means lie) for both the arithmetic and geometric means (Figure 7). This confirms that the arithmetic mean is substantially less precise than the geometric mean. The bootstrap confidence interval estimates tend to be a bit narrower than those with the normal approximation used in the figures above for the geometric mean (using formula 2ci). The widths are larger for earlier years than for later years, in contrast to Figure 6, because the means have not been divided by the overall means, which are much larger for early years than for later years. The low means are also the reason for the 2015 widths not being substantially larger than the previous widths, in contrast to both Figure 5 and Figure 6.

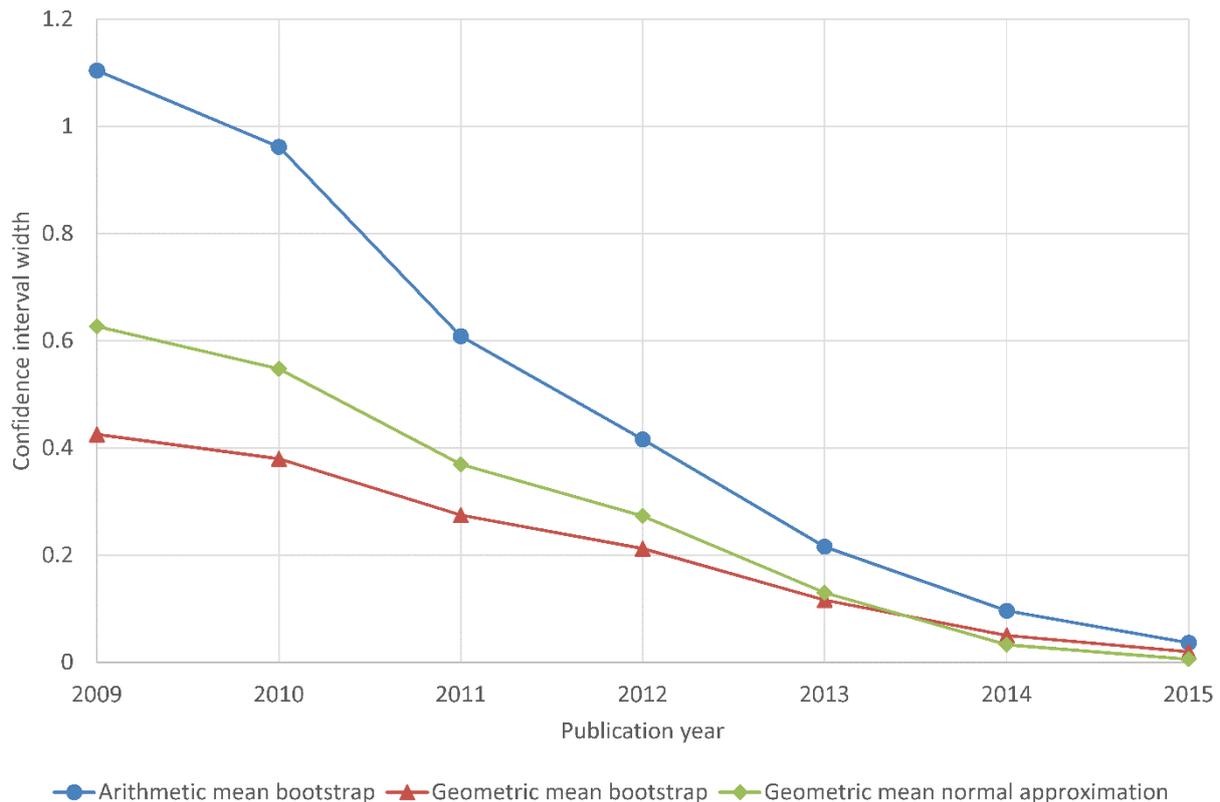

Figure 7. 95% confidence interval widths for the geometric or arithmetic citation mean for each subject. Each point in the graph is the median across the 26 subjects.

The 95% confidence intervals for the top 10% share for each country depend mainly upon the sample size (the number of articles for each country), and hence are approximately constant for each country, except for 2015 with a smaller number of articles (Figure 8). Countries tending to have smaller proportions in the top 10% tend to have narrower confidence intervals too. The width of the confidence intervals depend on the value of X because of the formula 4ci. For example, confidence intervals for 1% would be approximately a third (10/√11) as wide as confidence intervals for 10%. The widths are not directly comparable to those in Figures 5 to 7, however, because the measurements are in different units.

One way to compare confidence intervals between indicators (e.g., Figures 5 and 7) is to divide the confidence interval width by the range between the different countries. Ignoring the outliers of Russia and 2015, the typical range for the geometric mean is 0.6: from 0.8 to 1.4 (Figure 2), and the typical range for the proportion in the most cited 10% is 0.09: from 0.06 to 0.15 (Figure 4). Again ignoring the outliers of Russia and 2015, the typical confidence interval width for the geometric mean is 1.6% or 0.016 (Figure 6), and the typical confidence interval width for the proportion in the most cited 10% is 0.07 or 7% (Figure 8). Hence, for the geometric mean the typical range is 2.7 times (i.e., 0.6/0.22) as wide as the typical confidence interval width, whereas for the proportion in the most cited 10% the typical range is 1.3 times (i.e., 0.09/0.07) as wide as the typical confidence interval width. Thus, the population proportion estimate seems to be half as precise as the geometric mean. This is consistent with the geometric mean results being more stable over time (Figure 2) than the proportion results (Figure 4) in the sense that the lines are noticeably less jagged. Intuitively, also the proportion of the world's excellent articles written by a country seems to be less stable than the average citation count of all articles, if outliers are minimised with the geometric mean.

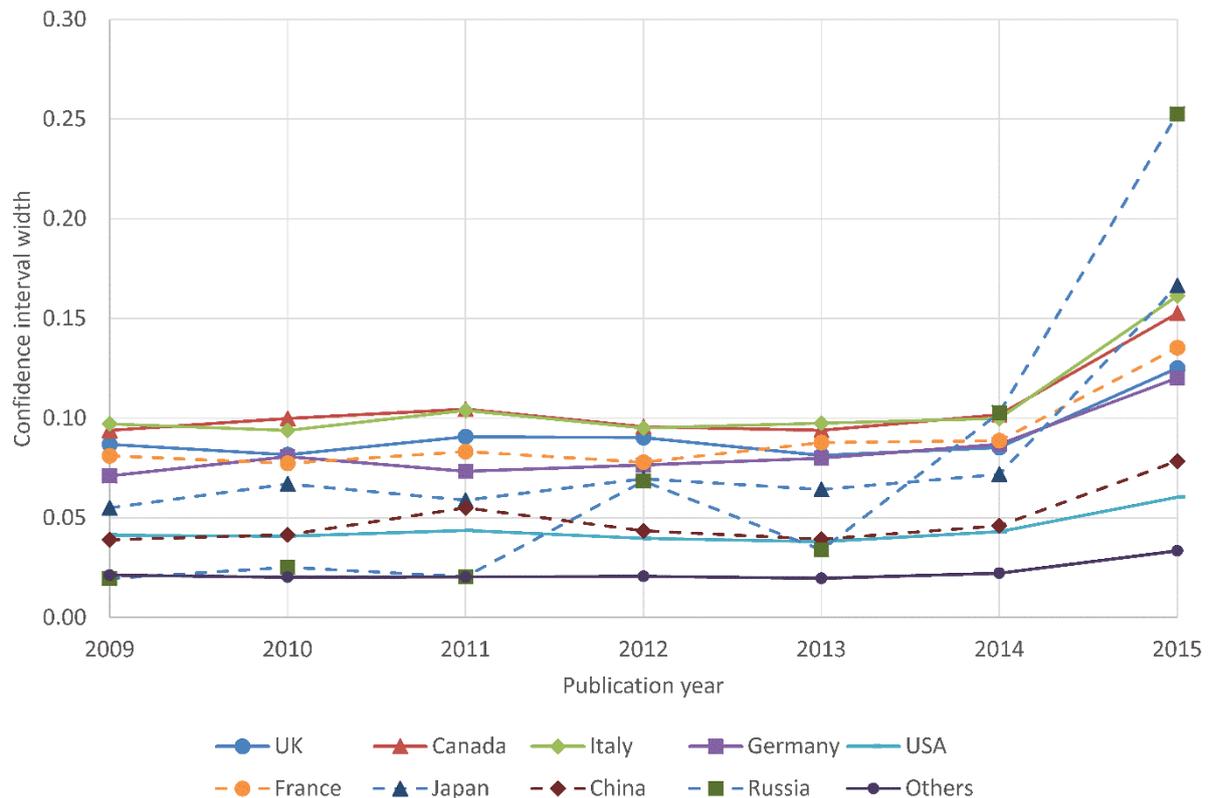

Figure 8. 95% confidence interval widths for the percentage of articles in the most cited 10% for each subject. Each point in the graph is the median across the 26 subjects.

## 5. Discussion and conclusions

The new methods introduced are an attempt at more precise indicators of citation impact for international comparisons between fields, although the results above have focused on aggregate results rather than the 26 individual subjects. The techniques used to answer the research questions have several limitations. The results may vary between subjects if some have considerably more or less varied citation counts for articles. The results for individual subjects are also likely to be substantially different with different field definitions or citation indexes. The methods would need to be modified if applied to simultaneously compare sets of articles from multiple subjects, such as articles funded by different funding agencies or funding streams. They also have the disadvantage that the geometric mean is much less well known than is the arithmetic mean, making interpretations of the results by policymakers, who are the end users, more difficult.

The results reveal some international trends about the subject area medians. From the most precise of the estimates (Figure 3), ignoring the imprecise data for 2015, it is clear that Italy's median national citation impact for the selected 26 subject areas has increased over time relative to the world average, whereas those of the UK, USA, Canada, France and China have decreased, and Russia performs substantially below the world average in terms of Scopus-indexed citations. These trends vary between subjects, however. For example, in Language and Linguistics, Italy's national citation impact decreased from 2009 to 2014, and Russia is approximately level overall with China for Computational Theory and Mathematics and for Animal Science and Zoology.

The two new methods introduced here for calculating national citation impact estimators for individual subject areas within different countries give results that are broadly comparable overall to the previous arithmetic mean method, giving them credibility as potential replacements. Although the linear regression model has the advantage that it could distinguish between the contributions of different nationality authors to international collaborative articles, it is less precise than the geometric mean and so is less useful for individual subjects. Nevertheless, even the confidence intervals for the national citation impact estimators using the geometric mean tend to be too wide to be able to robustly distinguish between most pairs of nations, despite median sample sizes above 5,000 for all years before 2015. The geometric mean is also more precise than the most cited 10% proportion indicator. In conclusion, although the geometric mean national citation impact estimators method is recommended for future international citation impact comparisons between fields, its values are not likely to be precise enough to distinguish between countries for individual subjects unless they have widely differing citation impacts within the subject. Thus the figures should be interpreted as rough guidelines about likely differences rather than robust evidence. Of course, human interpretation is also necessary to give context to the results, as always with citation indicators. This is because of the sources of systematic bias in the data, such as uneven coverage of national journal literatures.

The two new methods may also be useful for overall cross-subject international comparisons, although the citations would need to be normalised separately by field (using the new crown indicator approach: Waltman, van Eck, van Leeuwen, Visser, & van Raan, 2011a) before overall aggregate results could be calculated. The new methods would give more precise results but the added precision may not be worth the increased difficulty of interpretation for policy makers. Nevertheless, the additional precision may be essential for nations that do not have enough Scopus-indexed publications to get robust national impact indicators otherwise. The same applies to others that need to evaluate the impact of multidisciplinary sets of articles, such as research funding agencies.

Finally, for future work it would be interesting to assess the new methods for sets of individual universities rather than individual nations, with appropriate field normalised citations. Presumably they would work reasonably well for the largest institutions but give much less stable results for those that are small.